\documentclass[lettersize,journal]{IEEEtran}
\author{
    \IEEEauthorblockN{
        Jake McNaughton\textsuperscript{1},
        A.~H.~Abbas\textsuperscript{1,*},
        Ivan~S.~Maksymov\textsuperscript{1,2,$\dagger$}
    }
    \\
    \IEEEauthorblockA{
        \textsuperscript{1}Artificial Intelligence and Cyber Futures Institute,\\Charles Sturt University, Bathurst, NSW 2795, Australia. \textsuperscript{*}Email:~aborae@csu.edu.au
    }
    \\
    \IEEEauthorblockA{\textsuperscript{2}Seymour Research Laboratories, Seymour, VIC 3660, Australia. \textsuperscript{$\dagger$}Email:~ivan.maksymov@gmail.com}
}

\usepackage{amsmath,amsfonts}
\usepackage{algorithmic}
\usepackage{algorithm}
\usepackage{array}
\usepackage{hyperref}
\usepackage[caption=false,font=normalsize,labelfont=sf,textfont=sf]{subfig}
\usepackage{textcomp}
\usepackage{stfloats}
\usepackage{url}
\usepackage{verbatim}
\usepackage{graphicx}
\usepackage{cite}
\usepackage{color}
\usepackage{threeparttable}

\definecolor{my_green}{rgb}{0.55, 0.71, 0.0}

\begin{document}

\title{Full-Precision and Ternarised Neural Networks with Tunnel-Diode Activation Functions: Computing and Physics Perspectives}

\maketitle
\begin{abstract}
The mathematical complexity and high dimensionality of neural networks slow both training and deployment, demanding heavy computational resources. This has driven the search for alternative architectures built from novel components, including new activation functions. Taking a different approach from state-of-the-art neural and neuromorphic computational systems, we employ the current–voltage characteristic of a tunnel diode as a quantum physics-based activation function for deep networks. This tunnel-diode activation function (TDAF) outperforms standard activations in deep architectures, delivering lower loss and higher accuracy in both training and evaluation. We also highlight its promise for implementation in electronic hardware aimed at neuromorphic, ternarised and energy efficient AI systems. Speaking broadly, our work lays a solid foundation for a new bridge between machine learning, semiconductor electronics and quantum physics---bringing together quantum tunnelling, a phenomenon recognised in six Nobel Prizes (including the 2025 award), and contemporary AI research.
\end{abstract}

\section{Introduction}

\IEEEPARstart{T}{he} field of machine learning is evolving quickly, driven by growing demand for energy-efficient and biologically plausible models. Neural networks depend on activation functions \cite{Hay98}---nonlinear transformations applied to each neuron's output---that enable deep architectures to learn complex structure in data~\cite{Cyb89, Hay98, GooBook}. This nonlinearity is essential:~without it, a network collapses to a simple linear model and cannot capture the layered representations present in real-world signals~\cite{Hay98}. Yet widely used activations such as ReLU, sigmoid and tanh face well-known limitations in deep learning, including vanishing gradients and poor energy efficiency~\cite{Hay98, Kim17, GooBook}. These problems restrict scalability and slow learning in deep networks, underscoring the need for alternative activation mechanisms that deliver robust performance with lower computational and power costs~\cite{GooBook, Xu15, Ram17}.

Physics, particularly quantum mechanics, has profoundly influenced machine learning, inspiring the development of novel algorithms that exploit physical principles to enhance computational efficiency and reduce power consumption \cite{Mar20, Nak22, Mak23_review, Abb24_1}. A notable example is quantum reservoir computing \cite{Mar20_2, Dud23, Abb24}, a subclass of neuromorphic and unconventional computing systems \cite{Ada17} that employs the intrinsic dynamics of physical systems to enable energy-efficient machine learning solutions, particularly for predicting highly nonlinear and chaotic time series, as well as for pattern recognition tasks \cite{Dud23, Abb24}. Concurrently, spiking resonant tunnelling-diode neurons and neural networks~\cite{OPiwonka21, OPiwonka21_1, Don24, Owe25, Jac25} are gaining attention for their ability to reproduce key biological neuronal behaviours, including excitable spiking and refractoriness, while offering high operating speeds, energy efficiency and compact device footprints.
\begin{figure*}
    \centering
    \includegraphics[width=0.7\linewidth]{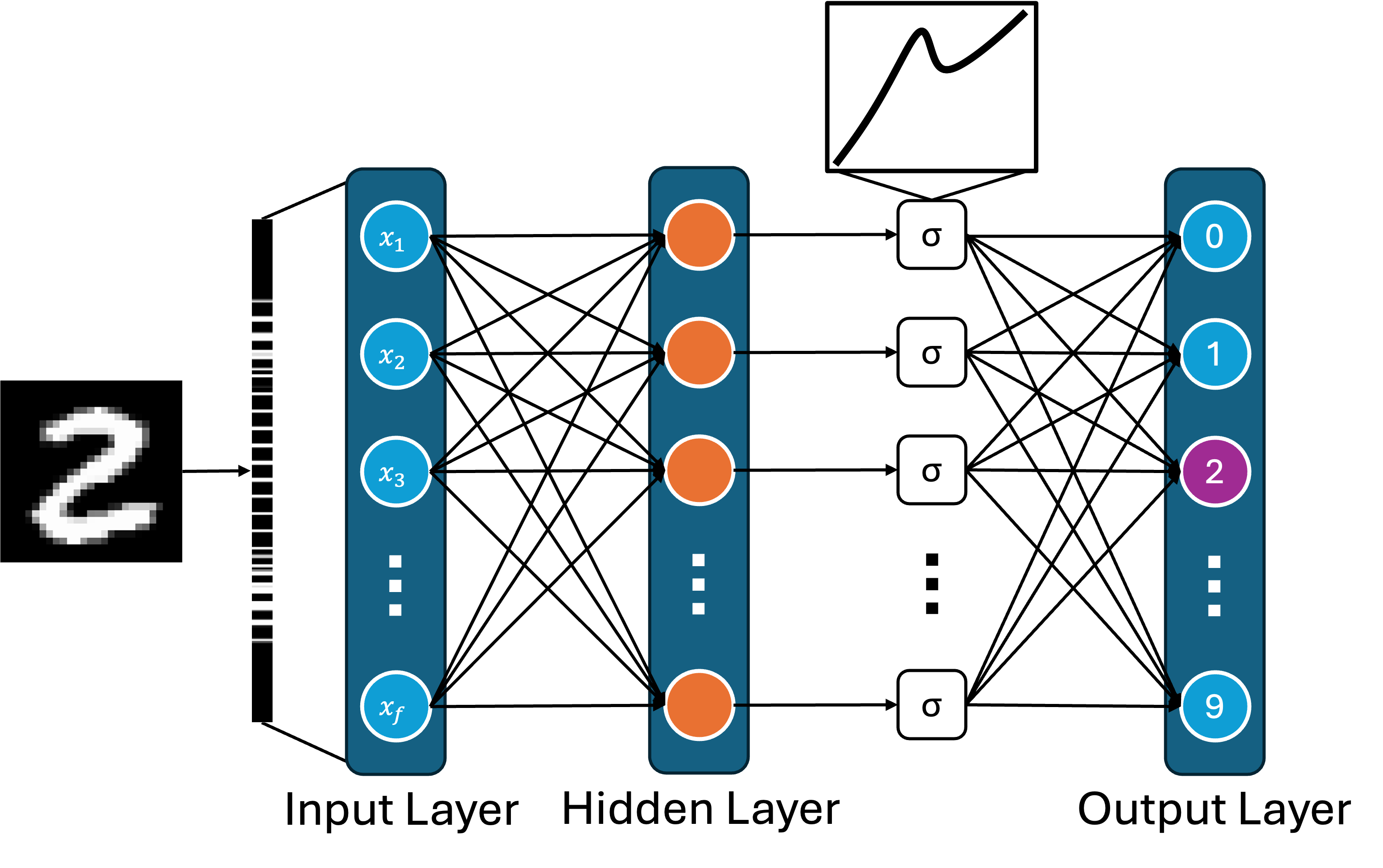}
    \caption{Architecture of a neural network using the current–voltage characteristic of a tunnel diode (inset) as a nonlinear activation function, referred to as TDAF in the main text.}
    \label{fig:nn}
\end{figure*}

In this paper, we take a fundamentally different path from neuromorphic spiking systems~\cite{OPiwonka21, OPiwonka21_1, Don24, Owe25, Jac25}---and, indeed, from any computational architecture that employs tunnel diodes or related quantum electronic devices~\cite{Lee22}. Instead, we use the current–voltage characteristic of a tunnel diode as a quantum-physics-based activation function for deep neural networks.

Although our work ultimately converges with ongoing efforts to develop more efficient neuromorphic neural network systems employing tunnel diodes~\cite{OPiwonka21, OPiwonka21_1, Don24, Owe25, Jac25}, its conceptual foundations arise from a completely different direction. In our case, the idea of employing quantum-tunnelling (QT) electron devices \cite{Esa58, Cha74} within a neural network originates in research on quantum cognition theory~\cite{Khr06, Atm10, Bus12, Pot22}, where quantum statistics has been shown to substantially outperform classical models~\cite{Gal_book} in explaining human cognitive behaviour, perception and decision-making~\cite{Mak24_APL, Mak24_illusions, Maks25}. 

Beyond the intriguing fact that QT, the defining physical effect underpinning tunnel diodes, offers a novel pathway for modelling the complexities of human cognition, QT has been employed in neuromorphic architectures that exploit negative differential resistance~\cite{Yil13, Ken24}. QT-based devices also provide substantially lower power consumption than traditional integrated circuits~\cite{Fen23} and possess properties essential for operation in harsh environments such as space~\cite{Rev78}. QT underlies the operation of scanning tunnelling microscopy~\cite{Bin87} and has been observed in quantum dots~\cite{Yil13}, which themselves constitute key building blocks in neuromorphic systems~\cite{Mar20_2}.

Recent advances in quantum computing further incorporate QT in quantum annealing and neuromorphic architectures~\cite{Abe21}, where tunnelling enables systems to traverse energy barriers and identify optimal solutions in combinatorial optimisation tasks~\cite{Abe21}. QT-based mechanisms have also been integrated into computational models inspired by brain-like architectures~\cite{Che24} and QT plays a recognised role in quantum biology, influencing enzymatic processes and energy transfer in photosynthesis~\cite{Geo19}.

The mathematical expressions governing the probability of QT through a potential barrier exhibit rich nonlinear properties \cite{Mak24_APL} and semiconductor diodes serve as a key element of NC systems \cite{Abb25}. Therefore, in this paper we propose and theoretically demonstrate that the current–voltage ($I–V$) characteristic of a tunnel diode can serve as a novel nonlinear activation function in neural networks (Fig.~\ref{fig:nn}). Unlike conventional activation functions, which are often heuristic and lack direct physical interpretations, the intrinsic nonlinearity of the tunnel diode arises from well-defined quantum mechanical principles \cite{Mes62, Sze_book}. We show that the neural network implementing the tunnel-diode activation function (TDAF) not only exhibits superior performance in terms of convergence, accuracy and expressiveness of the model but also introduces an energy-efficient approach to neuromorphic computing by extending prior relevant research work on neural networks with ternarised weights and activation functions~\cite{Par25, Jia25}.

Finally, our work establishes a solid foundation for a new bridge between machine learning, semiconductor electronics and quantum physics---uniting quantum tunnelling, a phenomenon recognised by six Nobel Prizes (including the 2025 award)~\cite{Mer02}, with contemporary AI research.

\section{Methods}
\subsection{Traditional Activation Functions}
The nonlinearity of activation functions enables neural networks to model complex patterns beyond what linear transformations alone can achieve \cite{GooBook}. Various activation functions have been developed, each with advantages and limitations \cite{GooBook, Ram17}.

For example, sigmoid $f(x) = \dfrac{1}{1 + e^{-x}}$ and hyperbolic tangent (tanh) $f(x) = \dfrac{e^x - e^{-x}}{e^x + e^{-x}}$ functions \cite{Luk09, Luk12, GooBook} compress inputs into a fixed range $(0, 1)$ and $(-1, 1)$, respectively. However, while useful for probability estimation and feature scaling, these functions suffer from saturation, where large or small inputs lead to near-zero gradients, hindering weight updates in multilayer networks \cite{GooBook}.

Therefore, one of the most widely used activation functions is the Rectified Linear Unit (ReLU), defined as $f(x) = \max(0, x)$ \cite{GooBook}. ReLU is computationally efficient and helps mitigate some of the vanishing gradient problems associated with sigmoid and tanh functions. However, ReLU is not fully immune to gradient-related issues, since neurons based on this function can become inactive when they consistently receive negative inputs, which prevents further learning \cite{GooBook}.

To address this limitation, variants such as Leaky ReLU and Parametric ReLU have been proposed, which introduce small negative slopes for negative inputs to maintain gradient flow and prevent inactive neurons \cite{Xu15}. Other functions such as Swish $f(x) = x \cdot \sigma(x)$, where $\sigma(x)$ is the sigmoid function, and GELU have been introduced to balance smoothness and gradient propagation \cite{Ram17}. Commonly used in transformer models, these activation functions offer improved expressiveness but remain not universally optimal \cite{Ram17}.

Thus, no single activation function is ideal for all tasks, as each has trade-offs in efficiency, stability and expressiveness \cite{Ram17}. The quest for a universal nonlinear function remains open, with researchers exploring biologically inspired models, adaptive activation mechanisms and quantum-inspired functions that could enhance learning dynamics across diverse architectures and applications \cite{Mar20_2, Mak24_APL}.
\begin{figure}
    \centering
    \includegraphics[width=\linewidth]{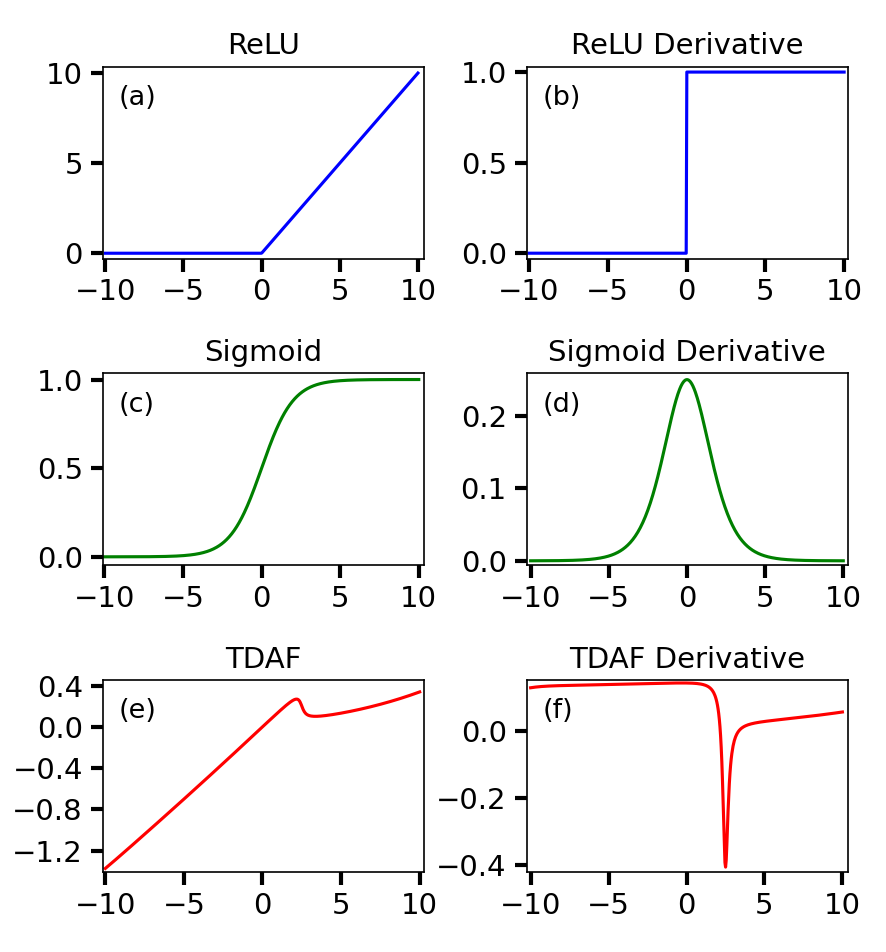}
    \caption{Two commonly used activation functions (ReLU and sigmoid) and their derivatives, alongside the proposed activation function, TDAF.}
    \label{fig:activations}
\end{figure}

\subsection{Tunnel-Diode Activation Function}
A tunnel (Esaki) diode is a heavily doped $p–n$ junction electron device whose $I-V$ characteristic contains a region of negative differential resistance arising from quantum-mechanical tunnelling through a narrow potential barrier~\cite{Esa58, Sze_book}. Heavy doping reduces the depletion width to only a few nanometres, aligning the $n$-region conduction band with the $p$-region valence band and enabling electrons to tunnel directly through the junction. Under zero or small forward bias this band alignment produces a sharp increase in current. As the voltage increases, band misalignment lowers the tunnelling probability and the current falls, giving rise to the characteristic negative differential resistance region.

Resonant-tunnelling diodes form a related class of devices in which quantum wells and double potential barriers enable resonant transport. Although structurally different, Esaki diodes and resonant-tunnelling diodes share the same physical basis:~voltage-dependent QT that yields a highly nonlinear transport response~\cite{Sze_book, Cha74, Sun98, Yan08}. While the present work employs a theoretical model originally developed for resonant-tunnelling diodes~\cite{OPiwonka21, OPiwonka21_1}, the discussion applies broadly to tunnel diodes. Crucially, a tunnel diode provides a strongly nonlinear, voltage-dependent effective resistance~\cite{Abb24_1}, making it well suited as a physics-based activation function.

Mathematically, the $I-V$ characteristic curve of a tunnel diode is a highly nonlinear function, primarily due to a negative differential resistance region, and it is given by
\begin{equation} \label{eq:1}
    I(V) = J_{1}(V)+J_{2}(V)\,,
\end{equation}
where 
\begin{equation} \label{eq:term1}
    J_{1}(V) = a\ln \left( \frac{1 + e^{\alpha + \eta V}}{1 + e^{\alpha-\eta V}} \right)  \times \left(\frac{\pi}{2} + \tan^{-1} \left( \frac{c - n_1V}{d} \right)\right)\, \nonumber
\end{equation}
and $J_{2}(V) = h \left(e^{\gamma V} - 1 \right)$.

The parameters are defined as $\alpha = \frac{q (b - c)}{k_B T}$, $\eta = \frac{q n_1}{k_B T}$ and $\gamma = \frac{q n_2}{k_B T}$, where $q$ is the electron charge, $k_B$ is Boltzmann's constant, $T$ is the temperature, $V$ is the voltage, and $a$, $b$, $c$, $d$, $n_1$ and $n_2$ are parameters characterising the physical system. In this work, we use $T = 300$\,K, $a = 0.0039$\,A, $b = 0.5$\,V, $c = 0.0874$\,V, $d = 0.0073$\,V, $n_{1} = 0.0352$, $n_{2} = 0.0031$ and $h = 0.0367$\,A, with all as per Ref.~\cite{OPiwonka21} except $b$, which has been scaled by a factor of $10$. These coefficients act as tunable parameters in the model and may be linked to relevant semiconductor physics quantities~\cite{Sch96}. 

Thus, Eq.~(\ref{eq:1}) provides an accurate fit to the $I-V$ characteristics of standard tunnel diodes because its fitting parameters are free to take values not strictly tied to direct physical interpretation~\cite{Sch96}. Other models of tunnel diode large-signal behaviour have also been proposed~\cite{Our23} and these can be substituted for Eq.~(\ref{eq:1}) when analysing specialised or non-conventional diode structures.

From both a physics and mathematics perspective, TDAF, as defined by Eq.~(\ref{eq:1}), can be separated into three distinct regions \cite{OPiwonka21}: two regions of positive differential resistance separated by a central region of negative differential resistance (approximately in the $x$-axis range from 1 to 3 in Fig.~\ref{fig:activations}e). The first term of Eq.~(\ref{eq:1}), $J_{1}(V)$, dominates at lower voltages, producing the positive differential resistance region at low voltage and the negative differential resistance region, but fails to capture the region of increasing current after the tunnelling region. The second term, $J_{2}(V)$, dominates at higher voltages and produces the second positive differential resistance region \cite{Sch96}.

When compared to ReLU and sigmoid (Fig.~\ref{fig:activations}a,~c), TDAF, from a machine learning perspective, exhibits two main regions:~a nominally weakly nonlinear region for negative inputs which corresponds to the first positive differential resistance region of the $I-V$ characteristic, and a nominally highly nonlinear region for positive inputs which corresponds to the negative differential resistance region and the second positive differential resistance region of the I-V characteristic. Upon comparing the derivative plots of ReLU and TDAF (Fig.~\ref{fig:activations}b,~f), it is reasonable to expect that these two functions will exhibit similar machine learning behaviour. Additionally, we argue that TDAF exhibits a greater degree of nonlinearity compared to ReLU and sigmoid activation functions, as demonstrated below.

\subsection{Conventional Full-Precision Neural Network Model}
Feedforward neural networks, the foundational computational model considered in this paper, typically consist of an input layer, $L$ hidden layers and an output layer \cite{Mur06}. The $l$th hidden layer, containing $N_{l}$ neurons, takes an input vector, $x^{(l-1)}$, from the previous layer and applies a linear transformation 
\begin{equation}
F: x^{(l-1)} \to W^{(l)}x^{(l-1)}+b^{(l)}\,,
\end{equation}
where $W^{(l)}\in M^{N_{l-1}\times N_{l}}$ and $b^{(l)}\in \mathbb{R}^{N_{l}}$ are the weight matrix and bias vector of layer $l$ that are applied to input data received from layer $l-1$. An activation function, $\sigma^{(l)}$ is then applied to the output to obtain the input to the next layer $x^{(l)} = \sigma^{(l)}\left(a^{(l)}\right)$, where $a^{(l)}=F\left(x^{(l-1)}\right)$ \cite{Caterini2018, smets2024}. The neural network $\mathcal{N}$ is then the composition of the functions comprising these layers (see Fig.~\ref{fig:nn} for an illustration)
\begin{equation}
    \mathcal{N} := F_{L+1}\circ\sigma_{L} \circ F_{L}\circ \dots \circ \sigma_{1}\circ F_{1}\,,
\end{equation}
where $F_{L+1}$ is the linear transformation performed by the output layer. The input layer does not perform any computations.

Thus, nonlinearity is introduced into the neural network through the activation functions $\sigma_{1},\dots,\sigma_{L+1}$, which, in the framework of an individual model, are typically all the same function, $\sigma_{i}=\sigma: \mathbb{R}\to \mathcal{R}, \quad \forall i$ \cite{Cyb89, Hay98, GooBook}. Feedforward neural networks employing nonlinear activation functions have been shown to be universal function approximators \cite{Hor98}. For many common activation functions, the range, $\mathcal{R}$, is $[0, 1]$ or $[-1, 1]$.

Importantly, provided the domains encompass $\mathbb{R}$, an activation function $\sigma$ can be replaced with another function $\sigma^\prime$, such as TDAF (see the inset in Fig.~\ref{fig:nn}), without changing other aspects of the structure of the neural network. However, the optimal ranges for the hyperparameters, including learning rate, batch size and number of epochs, may differ between two distinct activation functions \cite{Maks25}.

In supervised classification problems, the network is responsible for determining the class to which a given sample from the dataset belongs \cite{GooBook}. The dataset consists of samples belonging to $C$ different classes. Each sample is a pair $(x_{i}, y_{i})$, where $x_{i}$ is the input data and $y_{i}$ is the one-hot encoding of the class that the data belongs to (a vector of length $C$ with a $1$ at the index corresponding to the class and zeros elsewhere). The output layer of the neural network produces a vector $\hat{y}$ of length $C$ and the index of the maximum value in the output vector corresponds to the predicted class.
\begin{figure*}[t]
    \centering
    \includegraphics[width=0.9\linewidth]{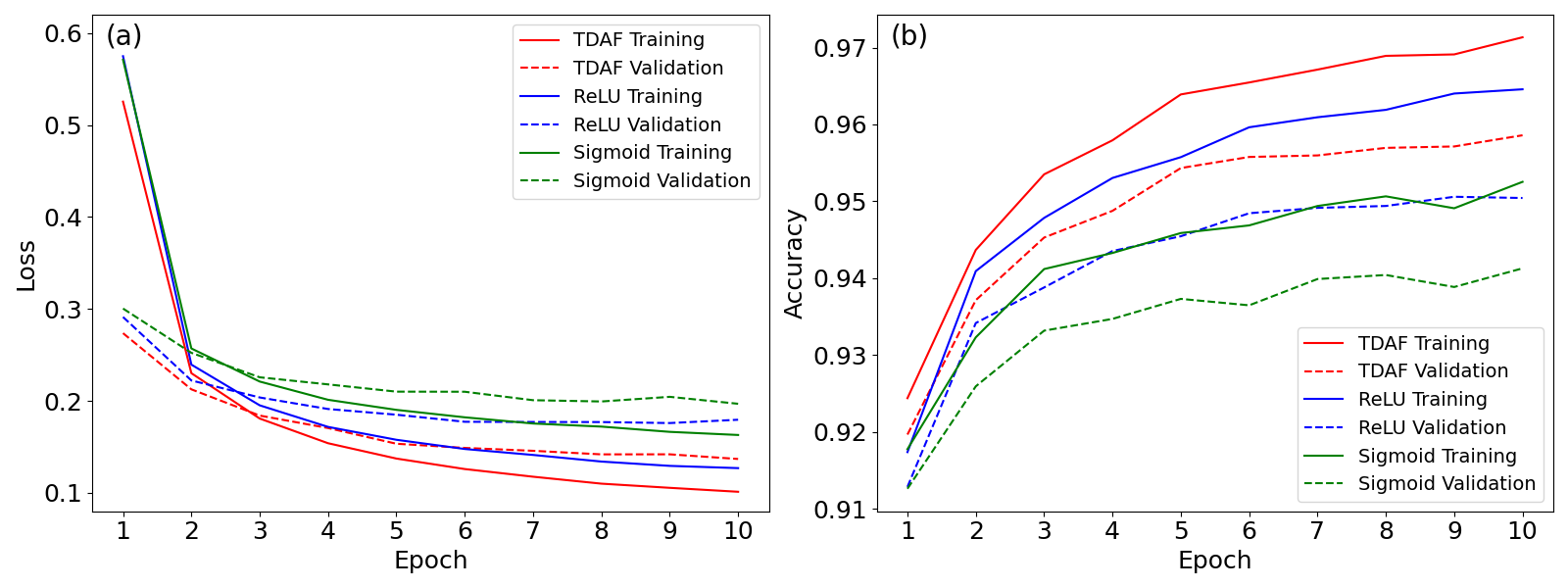}
    \caption{(a)~Loss curves obtained for the neural networks using ReLU, sigmoid and TDAF. (b)~Accuracy curves obtained for the neural networks using ReLU, sigmoid and TDAF. The initial accuracy values before the first epoch are not shown -- all were around 10\%.}
    \label{fig:curves}
\end{figure*}

In order for the network to improve on its predictions, a loss function $\mathcal{L}$ is required to quantify the performance. In classification tasks, cross-entropy is commonly used as the loss function \cite{GooBook}
\begin{equation}
    \mathcal{L}=-\sum_{c=1}^{C}y_{c}\log\left(\frac{e^{\hat{y}_c}}{\sum_{i=1}^{C}e^{\hat{y}_i}}\right)\,,
\end{equation}
where the argument of the logarithm can be interpreted as the predicted probability of sample belonging to class $c$ \cite{Ker18}, calculated from the output vector $\hat{y}$ as
\begin{equation}
    \hat{p_{c}} = \frac{e^{\hat{y}_c}}{\sum_{i=1}^{C}e^{\hat{y}_i}}\,.
\end{equation}
Optimising the loss function takes place through applying a gradient descent algorithm to the parameters $W^{(l)}$ and $b^{(l)}$ across all layers \cite{GooBook}. 

To optimise $\mathcal{L}$ with respect to the weights $W^{(l)}$ and biases $b^{(l)}$, gradient backpropagation is employed \cite{GooBook}. This involves computing the gradients of the loss function with respect to the network parameters. Using the chain rule, the gradient $\dfrac{\partial \mathcal{L}}{\partial W^{(l)}}$ is calculated as
\begin{equation}
 \frac{\partial \mathcal{L}}{\partial W^{(l)}} = \delta^{(l)} \cdot (x^{(l-1)})^{T}\,,
\end{equation}
where $\delta^{(l)}$ represents the error at layer $l$, propagated backward through the network
\begin{equation}
 \delta^{(l)} = \left((W^{(l+1)})^{T} \delta^{(l+1)} \right) \odot \sigma^{\prime}(a^{(l)})\,,
\end{equation}
where $(W^{(l+1)})^T$ is the transpose of the weight matrix from layer $l$ to $l+1$, $\odot$ is the element-wise multiplication operator and $\sigma^{\prime}(a^{(l)})$ is the derivative of the activation function at $a^{(l)}$. Then, the error at the output layer is computed as
\begin{equation}
\delta^{(L)} = \hat{y} - y\,.
\end{equation}

Using these gradients, the parameters are updated via gradient descent
\begin{equation}
 W^{(l)} \gets W^{(l)} - \eta \frac{\partial \mathcal{L}}{\partial W^{(l)}}\,,
 \end{equation}
 \begin{equation}
 b^{(l)} \gets b^{(l)} - \eta \frac{\partial \mathcal{L}}{\partial b^{(l)}}\,,
 \end{equation}
where $\eta$ is the learning rate parameter.

\subsection{Numerical Experiments and Datasets}
Figure \ref{fig:nn} illustrates the general feed-forward neural network architecture used in this paper, with the inset specifically showing how TDAF is incorporated into the model. Each network consists of three layers: an input layer, a single hidden layer and an output layer. In the first set of numerical experiments, three topologically identical neural networks were implemented in PyTorch, with the only distinction being the activation function applied after the hidden layer: ReLU, sigmoid or TDAF.

The standard MNIST handwritten digits dataset \cite{Den12} was used to benchmark the model. MNIST consists of 70,000 $28 \times 28$ images that are split into predefined training and test sets of 60,000 and 10,000 images, respectively. Consequently, the input layer received a $28\times 28$ image flattened into a one-dimensional vector of length 784. The flattened and normalised data served as the input into the first hidden layer, which was a fully connected layer consisting of 128 neurons. The output then passed through an activation function (ReLU, sigmoid or TDAF), before being sent through the output layer which consisted of 10 neurons, producing logits for classification.

Additionally, to prevent extreme numerical values from entering Eq.~(\ref{eq:1}), the input was clamped between $-10$ and $10$ (the choice of clamping limits depends on the specific data and task). The output was normalised between $0$ and $1$. The corresponding source code is available at \cite{Github}.

\section{Results}
The neural network architecture shown in Fig.~\ref{fig:nn} was trained through 10 epochs with a learning rate of 0.01 and a batch size of 256. The MNIST training sub-dataset was randomly split into 80\% training and 20\% validation, and the accuracy and loss values after each epoch were recorded for the training and validation sets. This training procedure was applied to the three topologically identical networks that utilised ReLU, sigmoid and TDAF, respectively. To account for variations caused by the random initialisation, the entire experiment was repeated 20 times. For each activation function, the means of the losses and accuracies after each epoch across the 20 repeats was recorded. All computations were performed on an Apple Macbook Pro M3 using an 8-core CPU.

\subsection{Loss and Accuracy Throughout Training}
Figure \ref{fig:curves}a shows the learning curves of the training and validation sets for the three different activation functions. We demonstrate that the network with the TDAF activation function achieves a lower loss at every epoch for both the training and validation sets. The loss of the TDAF network on the training set reached $0.1012$ compared to $0.1269$ for ReLU and $0.1630$ for sigmoid. On the validation set, TDAF reached a loss of $0.1369$, whereas ReLU reached $0.1796$ and sigmoid reached $0.1969$. Figure \ref{fig:curves}b shows the accuracy throughout training for both the training and validation sets. In this context, the TDAF network also shows superior performance across both the training and validation sets for all epochs.

\subsection{Accuracy on Test Dataset}
\begin{figure}[t]
    \centering
    \includegraphics[width=0.9\linewidth]{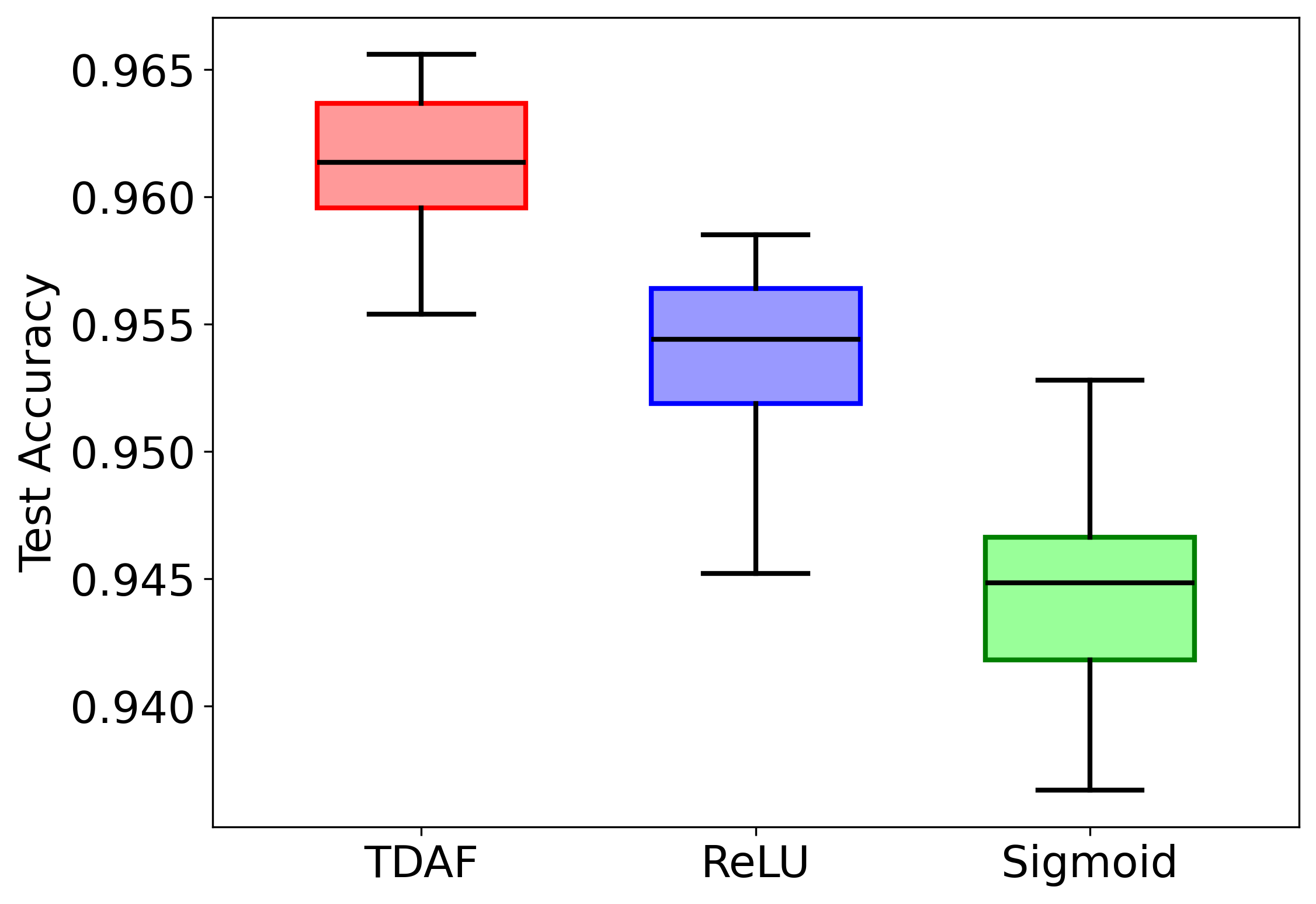}
    \caption{Box and whisker plot showing the median accuracy and interquartile range (IQR). The coloured boxes denote the value of IQR, with the straight line inside each box corresponding to the respective median value. The top and bottom straight lines located outside each box denote the high and low outliers, respectively.} 
    \label{fig:testsetresults}
\end{figure}

Figure \ref{fig:testsetresults} demonstrates the performance of the models on the test set after training, showing the median and interquartile range. TDAF achieves a mean accuracy of $96.14\%$ on the test set, compared to $95.32\%$ for ReLU and $94.44\%$ for sigmoid. This result is similar to the final accuracies of the validation set of $95.86\%$, $95.04\%$ and $94.13\%$ for TDAF, ReLU and sigmoid, respectively. Thus, outperforming the other activation functions based on the median and mean test accuracy, TDAF also exhibits a smaller interquartile range and overall range indicating more stable training dynamics. Additionally, TDAF outperforms both sigmoid and ReLU in terms of the loss and accuracy throughout training as well as the performance on the test set after training.
\begin{figure*}[!t]
\centering
\includegraphics[width=0.9\linewidth]{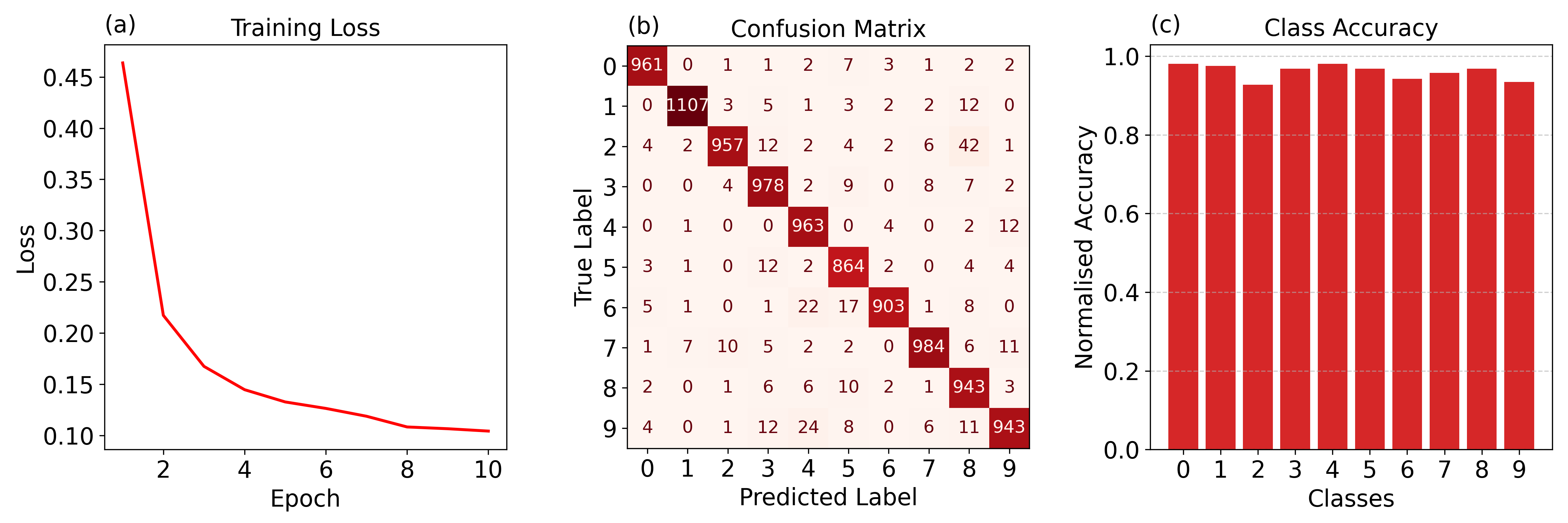}
\caption{Results from the TDAF neural network evaluated on the test set. (a) shows the loss throughout training, (b) shows the confusion matrix comparing the predicted labels to the true labels for each class, and (c) summarises the performance on each class.}
\label{fig_sim}
\end{figure*}

Since TDAF has demonstrated superior characteristics compared to other activation functions, we assess the performance of the TDAF-based neural network across the ten classes of the MNIST dataset. Figure \ref{fig_sim} presents the loss during training, the confusion matrix and the accuracy across different classes. The class accuracies range from $92.73\%$ for the digit 2 to $98.07\%$ for the digit 4. The most frequent misclassification was for the digits 2 being identified as a digit 8, occurring 42 times, an outcome that aligns with empirical observations \cite{Bal19}. It is noteworthy that the strong performance observed in this study, despite the simplicity of the feedforward neural network model, underscores the advantages conferred by TDAF, demonstrating its potential to enhance classification accuracy in neural networks with minimal architectural complexity.

\section{Comparison with a Quantised Neural Network}
While the class accuracies achieved in this work are already feasible, more advanced architectures, such as deep neural networks with extended hidden layers and convolutional neural networks, can reach even higher performance. In principle, the TDAF could be incorporated into these models; however, demonstrating state-of-the-art MNIST classification accuracy is not the objective of this article. Rather, our focus is on demonstrating that tunnel diodes, beyond their established applications in quantum-cognitive~\cite{Mak24_APL, Mak24_illusions, Maks25} and spiking~\cite{Lee22, Don24, Owe25} neural networks, offer a pathway towards inexpensive, low-power neuromorphic hardware~\cite{Abb24_1}. This naturally invites comparison with neural-network frameworks designed explicitly for efficient, resource-constrained computation, with quantised neural networks being the most widely adopted approach~\cite{Par25, Jia25}. 

Quantisation (not to be confused with the quantum-mechanical effects that underpin the operation of tunnel diodes) in neural networks is the process of reducing the numerical precision of weights, activations or both from standard 32-bit floating-point values to lower-precision formats such as 8-bit integers, ternary values $\{-1,\,0,\,+1\}$ and binary values~\cite{Par25, Jia25}. The aim of quantisation is to make the network cheaper and faster to run on a system with limited computational power without significantly degrading accuracy.

Lower-precision numbers reduce memory footprint, cut data-movement costs (a major source of energy consumption) and enable the use of simpler, more efficient arithmetic hardware~\cite{Ale17, Rut24}. This is essential for deploying deep learning models on edge devices, mobile processors, embedded systems and specialised accelerators where power, memory and computational resources are limited. Quantisation can also improve inference speed on general-purpose hardware, making it an important emerging machine learning technique~\cite{Ma24}.

To benchmark the performance of the TDAF-based neural network against a quantised model, we compare it with a customised ternary neural network~\cite{Github_ternary}. Our implementation incorporates a balanced combination of different techniques drawn from contemporary ternary neural network research, including modern weight-quantisation strategies~\cite{Den18, Li20_1, Li21} and teacher–student training approaches~\cite{Ale17}.

In ternary neural networks, real-valued weights are mapped onto the discrete set $\{-1,\,0,\,+1\}$ using a ternarisation operator. Let $w$ denote a real-valued weight and $\Delta > 0$ a threshold. The ternary weight $w^{t}$ is defined as
\begin{equation}
w^{t} = \operatorname{Tern}(w) =
\begin{cases}
+1, & w > \Delta, \\[4pt]
0, & |w| \le \Delta, \\[4pt]
-1, & w < -\Delta\,.
\end{cases}
\label{eq:ternarisation}
\end{equation}

Certain ternary models employ scaled ternarisation~\cite{Liu23} using a real-valued scaling factor $\alpha$, which captures the typical size of the significant weights, ignoring the small ones. This factor is introduced as
\begin{equation}
w^{t} = \alpha \cdot \operatorname{Tern}(w),
\end{equation}
where
\begin{equation}
\alpha = \frac{1}{|\mathcal{I}|} \sum_{i \in \mathcal{I}} |w_i|,
\qquad
\mathcal{I} = \{\, i : |w_i| > \Delta \,\}\,.
\end{equation}
However, in this work we modify this step and ternarise the weights directly as $w^{t} = \operatorname{Tern}(w)$~(i.e.~in our model, real-value scaling can yield adequate results; however, higher performance was achieved using fully ternarised weights).

We also use operator $\operatorname{Tern}$ as the activation function~\cite{Den18} as well as adopt a teacher–student training strategy, in which the forward activations and backpropagated gradients are computed using the ternarised weights $w^{t}$ and no bias term is included~\cite{Ale17}. In this setup, however, the underlying full-precision weights remain part of the optimisation loop and are updated in full precision throughout training. Moreover, the output layer of the network relied on raw scores with an approach inspired by squared hinge loss and its derivative~\cite{Den18}, enabling the final classification to be implemented directly via the {\tt argmax} decision function. Another advantage of using the balanced ternary system is that, during inference, the activation function reduces to a simple {\tt sign} operation~\cite{Ras16}.

However, because the ternarisation function in Eq.~\eqref{eq:ternarisation} is non-differentiable, gradient-based training requires an approximation to the derivative. This is typically achieved using the Straight-Through Estimator (STE) approach, which replaces the true derivative with an artificial mathematical expression during backpropagation~\cite{Ben13}. A general STE is expressed as
\begin{equation}
\frac{\partial w^{t}}{\partial w} \approx 1\,,
\label{eq:ste_general}
\end{equation}
allowing the gradient to `pass through' the quantisation step. We employ a more constrained version of the STE that limits gradient flow to a finite interval~\cite{Hub16}
\begin{equation}
\frac{\partial w^{t}}{\partial w} \approx
\begin{cases}
1, & |w| \le 1, \\[4pt]
0, & |w| > 1\,.
\end{cases}
\end{equation}

In Table~\ref{tab:model_accuracy}, we compare the final MNIST classification accuracies of the TDAF, ReLU and sigmoid full-precision models discussed above, along with the ternary model. We emphasise that our goal was not to fully optimise any of the models for maximum performance reported in the literature (see, e.g.,~\cite{Liu23}); therefore, the reported accuracies should be interpreted accordingly. We also note that the input to the ternary model consisted of ternarised MNIST images.

We observe that the TDAF and ternary models achieve comparable accuracies. However, while the ternary model restricts the use of full-precision arithmetic, since such operations would be prohibitively expensive when implemented in hardware such as field-programmable gate arrays (FPGAs)~\cite{Ale17}, the TDAF model, as presented here, retains full-precision activations in the mathematical framework used as a testbench. However, in practical hardware implementations, where TDAF units would correspond to physical tunnel-diode elements, the activation function would be realised directly in inexpensive, low-energy analogue circuitry, eliminating the need for full-precision digital computation~\cite{Abb24_1}.
\begin{figure}[t]
    \centering
    \includegraphics[width=0.7\linewidth]{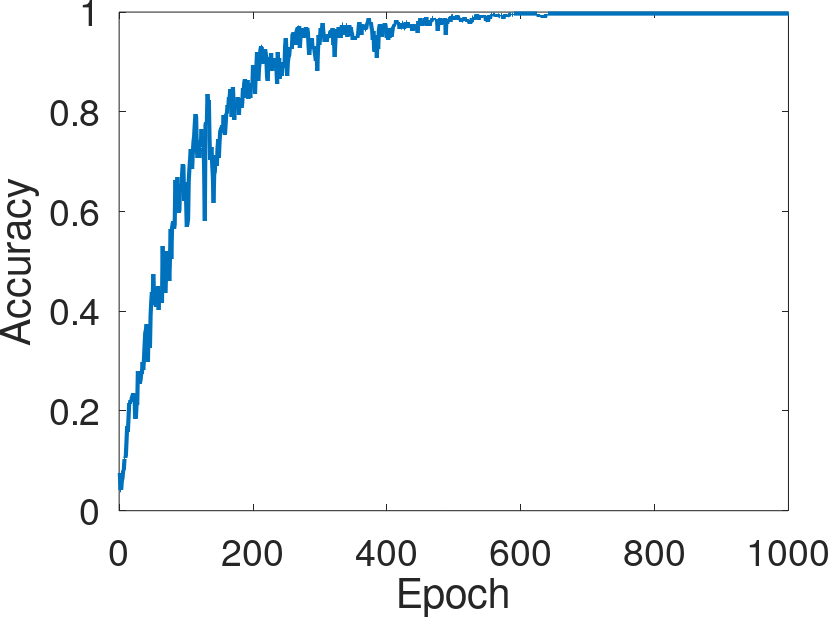}
    \caption{Convergence behaviour during training of the ternary neural network model used in this study.} 
    \label{fig:ternary_converge}
\end{figure}

It should be acknowledged that FPGAs are an established commercial technology but the use of analogue electronic devices in neuromorphic-style systems is still an emerging approach transitioning from laboratory prototypes to early industrial adoption. The TDAF model nevertheless can offer several potential advantages over the ternary model. First, implementing ternary representations on FPGAs requires additional workarounds, such as encoding trits using multi-bit binary patterns (e.g., 01, 10, etc. or similar schemes) to represent ternary values $\{-1,\,0,\,+1\}$, which introduces overhead compared with native binary hardware~\cite{Github_vhdl}. Second, as also noted in~\cite{Den18}, ternary models tend to converge more slowly than their full-precision counterparts (Fig.~\ref{fig:ternary_converge})---and therefore slower than the TDAF model, which in our experiments demonstrated faster convergence than both the ReLU- and sigmoid-based models. Third, as will be discussed in more detail below, inexpensive tunnel diodes can operate in harsh environments such as high altitude and space. However, commercial off-the-shelf FPGAs are not inherently radiation-hard and are vulnerable to single-event effects, which is especially critical for aerospace, high-altitude, nuclear and defence applications.
\begin{table}[t]
\centering
\begin{threeparttable}
\caption{Model Accuracy Comparison (MNIST test sub-dataset)}
\label{tab:model_accuracy}
\begin{tabular}{l c}
\hline
\textbf{Model} & \textbf{Final accuracy} \\
\hline
TDAF-DNN        & 95.86\% \\
ReLU-DNN        & 95.04\% \\
sigmoid-DNN     & 94.13\% \\
ternary-DNN     & 94.35\% \\
QT-ternary-DNN\tnote{a} & 84.00\% \\
\hline
\end{tabular}
\begin{tablenotes}
\footnotesize
\item[a] Hidden layers twice as small.
\end{tablenotes}
\end{threeparttable}
\end{table}

\section{TDAF: Origin of High Performance}
\subsection{Analysis of Nonlinear Properties of TDAF}
As discussed above, nonlinearity is essential for the operation of neural networks. In this section, we demonstrate that the increased nonlinearity of the TDAF plays a key role in its superior performance compared with traditional activation functions commonly used in machine learning. This discussion is warranted because earlier work on a tunnel-diode-based reservoir computing system~\cite{Abb25_1} has also demonstrated that increasing the nonlinearity of the signal processing in a tunnel diode improves the overall performance of the neural network.

Analysing the degree of nonlinearity of any signal is a non-trivial task, as no standard metrics exist and nonlinearity itself arises from diverse physical and mathematical processes. For example, one method to compare the nonlinearity of two signals is to analyse their Taylor series expansions and compare the magnitudes of the higher-order terms. However, the Taylor series is an approximation and calculating its coefficients can be non-trivial, especially for artificial functions such as ReLU.
\begin{figure}
    \centering
    \includegraphics[width=0.75\linewidth]{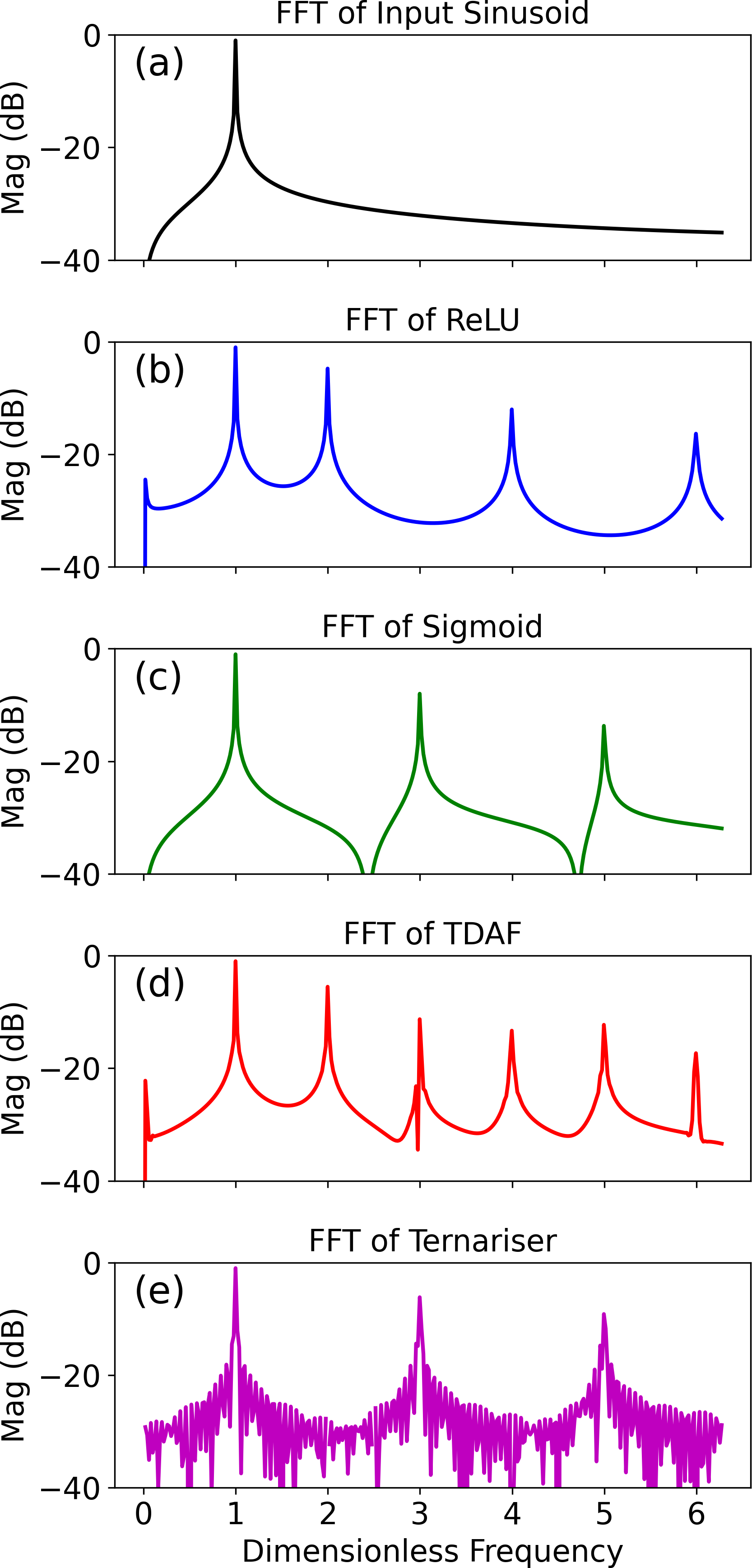}
    \caption{Fourier spectra of the outputs of (b)~ReLU, (c)~sigmoid and (d)~TDAF functions activated by a sinusoidal wave signal at a dimensionless frequency of 1. Panel~(a) shows the spectrum of the reference output produced by the linear identity function. The spectrum of the basic ternariser function is shown in Panel~(e).}
    \label{fig:TDAF_NL}
\end{figure}

Consequently, a previous study \cite{Mak19} suggested that two nonlinear physical phenomena can be compared by examining the Fourier spectra of a linear signal processed by the nonlinear systems of interest. Following this approach, we analyse the responses of the ReLU, sigmoid and TDAF activation functions to a purely sinusoidal wave signal at a dimensionless frequency of 1. For the sake of comparison, we also apply the sinusoidal signal to a linear identity function, producing a Fourier spectrum with a sole peak at 1. 

The choice of a simple sinusoidal input for analysing the activation function spectra is significant from both machine learning and neuromorphic computing perspectives. It has been shown that a conventional reservoir computing algorithm, which can in principle employ any standard neural activation function \cite{Luk09}, is mathematically equivalent to a model that combines the linear weights of the input dataset with its higher-order nonlinear functionals (quadratic, cubic, etc.)~\cite{Gau21}. The paper~\cite{Gau21} further demonstrates that computational efficiency can be enhanced by designing an algorithm, or using a suitably nonlinear theoretical or physical dynamical system, that sufficiently expands the input over a rich set of nonlinear functionals. This principle was confirmed experimentally using a fluid-mechanical dynamical system as a neuromorphic computer \cite{Mak24_dynamics}. Within this context, TDAF outperforms both ReLU and sigmoid activations because its Fourier spectrum exhibits all relevant nonlinear harmonics with comparatively higher amplitudes, thereby enabling more effective generation of nonlinear functionals of the input data and resulting in superior computational efficiency~\cite{Abb25_1}.

As shown in Fig.~\ref{fig:TDAF_NL}, the response of ReLU, sigmoid and TDAF to the sinusoidal waves results in the nonlinear generation of higher-order harmonics. According to \cite{Mak19}, the strength of the nonlinearity can be quantified using the magnitude of the harmonic peaks and the total number of harmonics produced by the nonlinear process. Focusing on the latter criterion, we can see that ReLU produces the peak at the second-harmonic frequency 2 and then at the fourth and the sixth harmonics. In turn, sigmoid produces the third and fifth peaks only. On the contrary, TDAF generates strong peaks at both odd and even harmonic frequencies. Within the theoretical framework used in this analysis \cite{Mak19}, this result suggests that the TDAF exhibits the strongest nonlinearity among the three analysed functions. We also observe that both TDAF and ReLU exhibit DC components in their respective spectra, which further supports our earlier assumption of their similar responses.

Interestingly, the similarity between the Fourier spectra of the sigmoid and ternariser activation functions arises because both introduce strong nonlinearity while preserving a smooth, bounded mapping of input amplitudes. The sigmoid smoothly compresses large inputs toward fixed asymptotic values, effectively saturating the signal and generating higher-order harmonics in its Fourier spectrum. A ternarised function, though discrete, approximates this behaviour by mapping small input variations near zero to a central state (0) and larger magnitudes to $\pm 1$, which retains the overall envelope and shape of the input signal. Consequently, both functions suppress extremely large deviations and concentrate the signal energy around the main frequency components of the original input, leading to a similar harmonic content and spectral distribution. In any case, within this analysis, the ternarisation function exhibits lower nonlinear transformation strength than TDAF.

\subsection{Probabilistic Ternarisation via Quantum Tunnelling}
Using an idealised physical model, we further suggest that the quantum-mechanical effect of QT, underlying tunnel-diode and TDAF operation, naturally enable these electronic devices to function as hardware quantisation units. The electric current through a tunnel diode can be expressed as an energy integral of the quantum transmission probability $T(E,V)$ weighted by the difference of the occupation functions on the two sides of the barrier (see~\cite{Ban89, Sch96, Geh06} and references therein). In general, we can write
\begin{equation}
I(V) \;\propto\; \int_{-\infty}^{\infty} T(E,V)\, \big[f_{p}(E-\mu_p)-f_{n}(E-\mu_n)\big] \, \mathrm{d}E\,,
\label{eq:I_general}
\end{equation}
where $f_{p},f_{n}$ are the Fermi--Dirac distributions on the $p$ and $n$
sides and $\mu_p-\mu_n = eV$ is the electrochemical potential difference
set by the applied bias $V$.

At low temperature, this reduces to an integral over the finite energy window where occupied states on one side overlap empty states on the other side
\begin{equation}
I(V) \;\propto\; \int_{\mu_n}^{\mu_p} T(E,V)\, \mathrm{d}E \,.
\label{eq:I_zeroT}
\end{equation}
Hence the voltage dependence of $I(V)$ is determined by how $T(E,V)$
varies with energy and how the bias window $[\mu_n,\mu_p]$ sweeps across
regions of high/low transmission.

For a simple rectangular barrier (height $V_0$, width $a$), the consideration of which suffices for application in the field of machine learning~\cite{Mak24_APL}, the transmission in the tunnelling regime ($E<V_0$) admits the approximate form
\begin{equation}
T(E)\;\approx\;\exp\!\big(-2\kappa a\big),\qquad
\kappa=\frac{\sqrt{2m(V_0-E)}}{\hbar},
\label{eq:T_approx}
\end{equation}
showing an exponential sensitivity to both $a$ and the barrier offset
$V_0-E$. Substituting \eqref{eq:T_approx} into \eqref{eq:I_zeroT} gives the
qualitative behaviour
\[
I(V)\;\propto\;\int_{\mu_n}^{\mu_p} \exp\!\big(-2a\sqrt{2m(V_0-E)}/\hbar\big)\,\mathrm{d}E\,.
\]

Two fundamental consequences follow from this relationship. First, regarding the peak current, as $V$ increases from zero, the bias window initially includes energies where $T(E)$ is large (resonant or high transmission),   so the integral in \eqref{eq:I_zeroT} and thus $I(V)$ increases. Concerning negative differential resistance, at larger $V$ the overlap of occupied and empty states moves out of the high-$T$ region and into energies where $T(E)$ is smaller; consequently the integral (and $I$) decreases with increasing $V$, producing the $N$-shaped $I-V$ curve characteristic of tunnel diodes. If resonant quasi-bound states exist inside the barrier~\cite{Geh06, Bos14}, their contribution may be modelled as a narrow resonance $S(E)$ so that the same integral formalism \eqref{eq:I_general} shows how resonances produce sharp peaks in $I(V)$.

Now we return to ternary machine learning models. A common alternative strategy for converting a continuous pre-activation $v$ into a ternary value $\{-1,0,+1\}$ is to use a probabilistic quantisation rule~\cite{Ale17}. In this approach, the ternary output is treated as a discrete random variable $W$ with probabilities
\begin{equation}
\begin{aligned}
P_{-1}(v) &= \Pr(W=-1\,|\,v),\\
P_{0}(v)  &= \Pr(W=0\,|\,v),\\
P_{+1}(v) &= \Pr(W=+1\,|\,v)\,.
\end{aligned}
\end{equation}
satisfying $P_{-1}(v)+P_{0}(v)+P_{+1}(v)=1$. One particular method decomposes the ternary choice into two nested Bernoulli decisions~\cite{Zhu17, Wan18, Li20_1, Bas21}. First, we need to decide whether the output is zero or non-zero
\begin{equation}
Z \sim \mathrm{Bernoulli}\!\left(p_{0}(v)\right), \qquad
\Pr(Z=1\,|\,v)=p_{0}(v)=P_{0}(v)\,.
\end{equation}
If $Z=1$, the output is set to $W=0$. If $Z=0$, the sign is sampled via a
second Bernoulli variable as
\begin{equation}
S \sim \mathrm{Bernoulli}\!\left(p_{+}(v)\right), \qquad
p_{+}(v)=\frac{P_{+1}(v)}{P_{-1}(v)+P_{+1}(v)}\,.
\end{equation}
The final ternary output is then
\begin{equation}
W =
\begin{cases}
-1, & Z=0,\, S=0, \\
+1, & Z=0,\, S=1, \\
\;\,0, & Z=1\,.
\end{cases}
\end{equation}

A simpler stochastic quantiser may use a noisy threshold around $v=0$~\cite{Ras16, Ale17}. Let $\eta$ be a zero-mean random variable (e.g.~Gaussian or uniform). Then
\begin{equation}
W = 
\begin{cases}
-1, & v + \eta < -\Delta, \\
\;\,0, & |v+\eta|\le \Delta, \\
+1, & v + \eta > +\Delta,
\end{cases}\,
\end{equation}
which implicitly defines probabilities
$P_{-1}(v)=\Pr(v+\eta<-\Delta)$,
$P_{0}(v)=\Pr(|v+\eta|\le\Delta)$ and
$P_{+1}(v)=\Pr(v+\eta>\Delta)$.

To employ TDAF in this context, we treat the neural pre-activation $v$ as controlling an effective particle energy $E(v)$ incident on a potential barrier~\cite{Mak24_APL}. For each trial, the particle (signal) may be reflected (ternary value $-1$), become trapped or be in temporary bound state~\cite{Bos14} (ternary value $0$) and transmit through the barrier (ternary value $+1$).

We can model these behaviours with three probabilities $P_{-1}(v)$, $P_{0}(v)$,
$P_{+1}(v)$ satisfying
\begin{equation}
    P_{-1}(v) + P_{0}(v) + P_{+1}(v) = 1.
\end{equation}

Form the physical point of view, for a rectangular barrier of height $V_{0}$ and width $a$, the quantum-mechanical transmission $T(E)$ is the following. For $E < V_{0}$ (tunnelling):
\begin{equation}
T(E)
=
\left[
1
+
\frac{V_{0}^{2}}{4E(V_{0}-E)}
\sinh^{2}(\kappa a)
\right]^{-1}\,,
\end{equation}
where $\kappa=\dfrac{\sqrt{2m(V_{0}-E)}}{\hbar}$. For $E > V_{0}$ (above-barrier):
\begin{equation}
T(E)
=
\left[
1
+
\frac{V_{0}^{2}}{4E(E-V_{0})}
\sin^{2}(ka)
\right]^{-1}\,,
\end{equation}
where $k=\dfrac{\sqrt{2m(E-V_{0})}}{\hbar}$.

We can set
\begin{equation}
P_{+1}(v) = T(E(v)),
\end{equation}
and model a temporarily bound state (`trapped') probability $S(E)$ as a resonance-like behaviour
\begin{equation}
S(E)
=
s_{0}\exp\!\left(
-\frac{(E-E_{r})^{2}}{2\sigma^{2}}
\right),
\end{equation}
where $E_{r}$ is a resonant energy and parameter $\sigma$ controls the width of the resonance lineshape. Renormalisation gives
\begin{equation}
P_{+1}=T, \qquad
P_{0}=S, \qquad
P_{-1}=1-T-S.
\end{equation}

To interface the dimensionless neural pre-activation $v$ with the quantum-mechanical tunnelling model, we can also map $v$ to an effective carrier energy using the transformation
\begin{equation}
E(v)=\alpha v + \beta\,.
\end{equation}

On physical grounds and based on machine learning parameter fine-tuning, the dimensionless parameters used in the QT ternarisation function are the following. Parameter $\alpha = 1.0$ (this is the scaling factor that controls how strongly the neural pre-activation $v$ influences the effective energy $E(v)$; larger $\alpha$ produce sharper transitions between ternary states). The offset $\beta = 1.5$ (it sets the reference energy level, determining the alignment of $E(v)$ with respect to the tunnelling barrier). The barrier height $V_0 = 1.51$ and width $a = 1.0$ define the rectangular potential barrier, governing the tunnelling probability according to quantum mechanics. Parameters $m = 1.0$ and $\hbar = 1.0$ are the effective particle mass and reduced Planck constant. Finally, $s_0 = 0.1$, $E_r = 1.5$ and $\sigma = 0.2$ characterise the resonance-like trapped state: $s_0$ sets the peak probability of being temporarily captured, $E_r$ is the resonant quasi-bound energy and $\sigma$ controls the width of the resonant peak.

We visualise and compare different representations of neural network weights for selected neurons from the first hidden layer of the network, revealing the effect of ternarisation on the weights. Specifically, full-precision weight vectors $W_1$ are reshaped into two-dimensional matrices and visually displayed alongside their conventionally balanced ternarised counterparts $btW_1$ and QT-ternarised versions $qtW_1$. For each selected neuron, the three representations are plotted side by side using greyscale colour maps, allowing a qualitative assessment of the sparsity and quantisation patterns induced by the ternarisation procedures.

In parallel, correlation coefficients between the full-precision and conventionally ternarised weights (approximately 0.85 on average across the neuron population), as well as between full-precision and QT ternarised weights (approximately 0.8), are computed, providing a quantitative measure of how closely the ternarised weights preserve the structure of the full-precision weights.

Obtaining correlation coefficients of this level suggests a moderate degree of linear association, with positive values in one weight distribution generally corresponding to positive values in the other. Thus, both ternarisation approaches moderately reduce the weight resolution and adequately retain structural features of the full-precision weights. In fact, as shown in Table~\ref{tab:model_accuracy}, a QT-ternarised neural network, with a deliberately reduced number of hidden layers by a factor of two to improve computational speed, confidently reaches 84\% classification accuracy.
\begin{figure}
    \centering
    \includegraphics[width=0.999\linewidth]{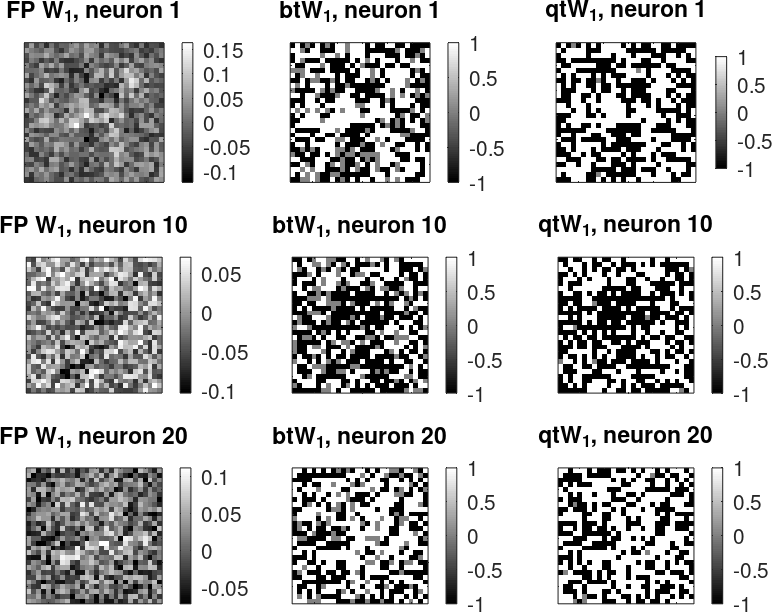}
    \caption{Visual comparison of weight representations for selected neurons in the first hidden layer. Full-precision weight vectors $W_1$ are reshaped into two-dimensional matrices and displayed alongside their conventionally balanced ternarised ($btW_1$) and QT-ternarised ($qtW_1$) counterparts. For each neuron, the three representations are shown.}
    \label{fig:neural_population}
\end{figure}

Finally, we note that although the software implementation of the QT-based ternarisation formulas may exhibit reduced computational performance relative to conventional ternarisation, the method is well suited for efficient hardware implementation. In analogy with FPGA-based realisations of standard ternarisation, the QT approach can be realised as a compact and low-cost electronic circuit employing tunnel diodes.

This raises the question of whether tunnel diodes can be integrated with FPGAs. Although these two technologies originate from distinct engineering traditions, their integration is not only feasible but technically advantageous for hybrid analogue–digital computing. At the practical level, tunnel diodes can be incorporated as discrete components on a printed circuit board and interfaced with an FPGA through analogue front-end circuitry comprising biasing networks, ADCs and DACs. In this configuration, the tunnel diode provides the high-speed nonlinear transformation required for the operation of the neural network, whereas the FPGA supplies reconfigurable digital control, state updating and post-processing. More sophisticated architectures could embed the tunnel diode within mixed-signal subsystems, allowing the FPGA to dynamically modulate its operating point and gain access to its nonlinear response with high temporal precision. While monolithic co-integration of tunnel diodes with CMOS, such as that used in commercial FPGA fabrication processes, remains limited to proof-of-concept demonstrations due to material and process incompatibilities, research prototypes based on resonant tunnelling diodes integrated with silicon electronics indicate that such hybrid platforms are technologically plausible~\cite{Sud04, Nag23}.

\section{General discussion} \label{disc}
The successful adaptation of the characteristic $I-V$ curve of a tunnel diode, and the quantum-tunnelling mechanism that underpins it, as an activation function for neural networks raises interesting questions about the underlying reasons for its improved performance and its potential future applications. TDAF exhibited superior training dynamics, stability and accuracy on the MNIST dataset compared with the sigmoid and ReLU activation functions. Alongside demonstrating the ability of TDAF to function as a quantiser for ternary neural network models, this suggests that TDAF produces a more efficient computing landscape compared with traditional machine learning activation functions.

The mechanism behind the superior performance of TDAF can potentially be explained by exploring the physical phenomena present in tunnel diodes---in particular, the effect of QT. The physically nonlinear (non-monotonic) nature of the TDAF, created by two positive differential resistance regions separated by a region of negative differential resistance, allows for efficient exploration of the local loss landscape, which may contribute to increased stability in training whilst exploring a broader area of the local optimisation landscape.

In practice, a TDAF-based neural network can be implemented using readily available commercial tunnelling \cite{Esa58} and resonant tunnelling \cite{Cha74} diodes. It is noteworthy that systems based on tunnelling diodes consume significantly less power than traditional integrated electronic circuits \cite{Fen23}, presenting an opportunity for developing semiconductor chips optimised for power-constrained applications, such as autonomous vehicles, where onboard energy resources are highly limited \cite{Abb24_1}. Furthermore, neuromorphic computing schemes that exploit negative differential resistance, a key characteristic of tunnelling diodes, have already been proposed \cite{Yil13, Ken24}, suggesting that TDAF functionality could be integrated into existing systems.

Alternatively, a TDAF can be implemented using scanning tunnelling microscopy instrumentation \cite{Bin87}. Amplifiers used in this kind of microscopy operate at microwave frequencies, providing high signal-to-noise ratios and enabling differential resistance spectroscopy measurements \cite{Bas18}, which can be used to compute the derivative of TDAF. In low-frequency designs, this derivative can also be obtained using simple analog electronic circuits \cite{Ulm20}. Furthermore, QT of individual electrons has been experimentally observed in quantum dots \cite{Yil13}, which serve as fundamental building blocks for quantum NC systems and can therefore be integrated with them \cite{Mar20_2}.

Artificial intelligence will be pivotal in future space exploration, but significant technological challenges must be addressed, particularly in energy consumption and thermal management \cite{Lv24}. Modern AI systems rely on efficient cooling to dissipate heat from graphics processing units, yet the vacuum of space renders conventional air-based cooling methods ineffective \cite{Lv24}. This makes heat dissipation a critical concern for spacecraft, satellites and space exploration vehicles, necessitating innovative cooling solutions to ensure the reliable operation of sensitive electronics in extreme environments \cite{NASA}.

Tunnel diodes, known for their high-speed operation, low noise and resilience in extreme conditions, are widely used in space applications such as high-frequency oscillators, amplifiers and satellite transceivers \cite{Rev78}. Their high radiation hardness, particularly in materials like gallium arsenide (GaAs) or indium phosphide (InP), ensures reliability in space missions, while their exceptional performance at cryogenic temperatures makes them valuable for astrophysics instruments \cite{Rev78}. This highlights the potential of TDAF as a transformative solution to replace the expensive, high-power-consuming and less reliable electronic systems currently used in ground-based AI.

\section{Conclusion}
We introduced a novel activation function, the tunnel-diode activation function (TDAF), derived from the current–voltage ($I-V$) characteristic of a tunnel diode. TDAF demonstrated superior performance compared to two widely used ReLU and sigmoid activation functions, achieving higher accuracy and lower loss during both training and evaluation. Furthermore, we demonstrated that TDAF can be used in ternary neural network models.

We suggested that the improved performance of TDAF can be attributed to the QT effect that underpins the operation of tunnel diodes. This effect introduces greater nonlinearity and expands data into a broader set of nonlinear functionals, enabling machine learning algorithms to process information more efficiently. This physically grounded approach opens new avenues for integrating quantum-inspired mechanisms into machine learning models, bridging the gap between AI and emerging neuromorphic computing platforms.

Future research will focus on further theoretical exploration of the QT-based nonlinearity of TDAF and its implications for network dynamics. Yet, practical implementation of TDAF in hardware-based neuromorphic circuits will be pursued, potentially enabling the development of energy-efficient, high-speed AI systems capable of operating in environments unsuitable for conventional or qubit-based quantum computing hardware.

Comparing the performance of TDAF implementations across different neural network architectures, such as recurrent and convolutional neural networks, reservoir computing systems and transformer-based models, will further test the robustness of the activation function and establish which types of tasks TDAF performs best. Moreover, studying how activation functions affect the loss landscape will provide insight into whether other physically inspired functions may offer advantages for machine learning applications.

As previously mentioned, our results indicate that implementing tunnel diode circuits, which take input data encoded as currents and process it analogously to a feedforward neural network, may yield promising outcomes. Pursuing this research direction could lead to cost-effective AI systems with improved computational efficiency, bridging the gap between quantum-inspired neuromorphic computing systems and practical machine learning implementations.

\section*{Data availability}
The codes and the datasets used in this research are publicly available and can be accessed at the following links:~\cite{Github, Github_ternary}.
\section*{Declarations}
The authors declare that there is no conflict of interest regarding the publication of this paper.

\bibliographystyle{IEEEtran}
\bibliography{references}

\end{document}